\documentclass[aps,prl,onecolumn,superscriptaddress,10pt,floatfix,amsmath,amssymb, showkeys]{revtex4-2}

\usepackage{graphicx}
\graphicspath{{images/}}
\usepackage{xcolor}
\usepackage{bm}
\usepackage{hyperref}
\usepackage[normalem]{ulem}
\usepackage{siunitx}
\DeclareSIUnit{\sample}{s}
\DeclareSIUnit{\bitpersecond}{bps}
\DeclareSIUnit{\decibel}{dB}
\usepackage{physics2}
\usepackage{booktabs}

\begin{document}

\title{Developing a photon-number-resolving detection chain for quantum communication protocols involving mesoscopic states of light}

\date{\today}

\author{Alex Pozzoli\textsuperscript{\P}}
\email{apozzoli4@uninsubria.it}
\thanks{$\,$\newline\noindent\hspace*{-8pt}\textsuperscript{\P} These authors contributed equally to this work.}
\affiliation{Como Lake Institute of Photonics, Department of Science and High Technology,
University of Insubria, Via Valleggio 11, 22100 Como, Italy}

\author{Stefano Carsi\textsuperscript{\P}}
\email{s.carsi@nuclearinstruments.eu}
\affiliation{Nuclear Instruments SRL, Via Lecco 16, 22045 Lambrugo, Italy}

\author{Andrea Abba}
\email{abba@nuclearinstruments.eu}
\affiliation{Nuclear Instruments SRL, Via Lecco 16, 22045 Lambrugo, Italy}

\author{Alessia Allevi}
\email{alessia.allevi@uninsubria.it}
\affiliation{Como Lake Institute of Photonics, Department of Science and High Technology,
University of Insubria, Via Valleggio 11, 22100 Como, Italy}
  
\begin{abstract}
\noindent We present the characterization of a photon-number-resolving detection chain based on Silicon photomultipliers (SiPM) coupled to a \SI{14}{\bit}, \SI{1}{\giga\sample\per\second} digital acquisition system embedding an FPGA-based signal processing pipeline that performs real-time baseline subtraction, digital deconvolution, and charge integration. Three SiPM models manufactured by Hamamatsu are tested and compared in the mesoscopic intensity regime using both classical coherent states and quantum twin-beam states, enabling a systematic investigation of the effects of pixel pitch, pile-up, and photon detection efficiency on the detector performance.
\end{abstract}


\maketitle

\section{Introduction}\label{sec1}
\noindent The realization of secure quantum communication protocols demands both fast light sources and acquisition devices \cite{usenko}. Currently laser systems employed in these scenarios can operate at tens of GHz \cite{paraiso}. 
Nowadays, continuous-variable quantum key distribution
offers high key rates over short distances, achieving rates over \SI{1}{\giga\bitpersecond} at 5-\SI{10}{\kilo\meter} and roughly 1-\SI{10}{\mega\bitpersecond} at \SI{25}{\kilo\meter}. Discrete-variable quantum key distribution typically operates at lower, more stable rates, ranging from a few kilobits to several megabits per second, but it is superior for long-distance (over \SI{100}{\kilo\meter}) and higher channel loss scenarios \cite{wang, pirandola}.
In discrete-variable receivers, the detection stage is based on single-photon detectors, usually single-photon avalanche photodiodes or superconducting nanowire single-photon detectors, whose signal processing has reached high optimization levels, especially in terms of speed and efficiency \cite{campbell,grunenfelder,zhang}. Reversely, continuous-variable receivers are based on optical homodyne detection \cite{roumestan}.\\
A different approach in communication protocols would be obtained by means of photon-number-resolving (PNR) detectors, which are able to discriminate the number of photons contained in each laser shot \cite{eaton,lita,schmidt}. 
This capability allows the exploitation of the number of photons as a degree of freedom in which to encode information. Moreover, PNR detectors provide the possibility to exploit mesoscopic states of light, which can be more robust against losses and noise in communication protocols. As an example, quite recently, some of us have presented novel communication protocols, in which the encoding of information involves the photon-number statistical properties of light, while the security level is ensured by the nonclassical correlations of mesoscopic twin-beam (TWB) states \cite{razzoli2025hybrid}.
Depending on the specific kind of detector that is exploited, a proper electronic treatment of the signal output should be designed since a standard and general solution still does not exist.\\
This is the case of Silicon photomultipliers (SiPMs) \cite{akindinov,ramilli}, which are PNR detectors, extensively exploited in particle physics experiments, that have received growing attention in the context of quantum technologies in the last decade \cite{chesi19}. Indeed, they can be operated at room temperature, are small, portable, and cheap. 
All these features make them particularly intriguing in specific environments. For instance, thanks to their high sensitivity in the blue spectral range, they could be useful for free-space underwater quantum communication \cite{allevi2022towards}. Moreover, given their compact dimensions, their robustness against magnetic fields and their operation at room temperature, they could be used for communication between satellites, where losses are dominated by diffraction and not by absorption. Furthermore, the compactness of SiPMs makes them suitable not only in direct detection schemes, but also in more complex ones. For instance, this is the case of the weak-field receiver \cite{donati,sanvito2024assessing,4states}, namely an interferometric scheme in which the signal is mixed with a mesoscopic local oscillator and the two outputs of the interferometer are detected by PNR detectors and suitably processed. This hybrid scheme allows the reconstruction of both particle-like and wave-like properties of light, working as a bridge between direct detection and homodyne detection. 
Specifically, it could be more advantageous than standard homodyne detection for the encoding of information in a continuous modulation \cite{cattaneo} or quasi-continuous one, such as in the case of quadrature amplitude modulation or amplitude-phase-shift keying, both requiring joint modulation of the amplitude and phase of optical coherent states \cite{hirano2017,ghorai2019}.\\
However, the output signal of SiPMs should be properly processed in order to decrease the effect of their drawbacks, especially dark counts \cite{dinu} and optical cross-talk \cite{nagy}. The reduction of them is typically achieved by restricting the integration to the fast charge component of the SiPM output, which corresponds to the first few nanoseconds of the signal, thereby suppressing both the contribution of dark counts, whose rate is uncorrelated with the signal, and that of afterpulses, which occur on a longer timescale and thus fall outside the integration window \cite{seifert}. In practice, this requires the acquisition system to resolve and integrate a signal window as short as a few nanoseconds, which in turn demands a fast, low-noise digitization front-end. Another important parameter to control is the acquisition speed. The maximum limit is given by the dead time of the cells forming SiPM detectors, which could be as short as 5-\SI{10}{\nano\second}, 
thus allowing a maximum acquisition rate of 100-\SI{200}{\mega\hertz}. Actually, faster solutions based on active quenching strategies are also possible, but in that case the PNR capability is greatly reduced to only a few photons \cite{lin}. In contrast, there could be situations in which slower acquisition rates are needed due to the amplification stage employed \cite{cassina21}.\\ 
To meet all these requirements, in our work we exploit a \SI{14}{\bit} digital charge integrator system combined to three different models of SiPMs. The device, embedded with an FPGA-based firmware, performs the digital deconvolution of the signal, compensates for the baseline level, and is also able to provide the pulse height spectrum in real time. 
The characterization of the complete detection chain is obtained by exploiting both classical and quantum states of light in the so-called mesoscopic intensity regime, in which each light pulse can contain several photons.
In particular, we test the minimum and maximum number of photons that can be properly detected and resolved by calculating different kinds of parameters, such as the visibility and the figure of merit \cite{brooks,recker}. Moreover, we also reconstruct the statistical properties of the employed light to test the accuracy of the obtained results. Specifically, we investigate the Fano factor, the correlation coefficient and the noise reduction factor in terms of numbers of detected photons \cite{PLA22}. The comparison among three different models of SiPMs also provides the possibility to study the effect of pile-up and quantum efficiency. The performed analysis highlights the versatility of the acquisition system and its usefulness in the context of quantum communication, as it guarantees a good dynamic range as well as a fast enough signal processing.
\section{Methods}\label{sec2}
\subsection{The detection chain} \label{th:det}
\noindent The detection chain tested in this work is based on different models of SiPMs, an amplification stage and a digital acquisition system also including a real-time data analysis stage. All measurements were taken in a temperature-controlled environment, and any thermal drifts were reduced to negligible levels by allowing all electronic devices to reach thermal equilibrium.\\
As for the detectors, we examine three different SiPM models, all manufactured by Hamamatsu, in order to make pair-by-pair comparisons. As extensively explained in Ref.~\cite{chesi19bis}, a SiPM is a matrix of cells operated in the Geiger-M$\rm \ddot{u}$ller regime with a common output. In general, such detectors are characterized by some drawbacks that should be kept under control \cite{chesi19bis}. First of all, they exhibit a non-negligible dark count rate (the typical values are tens to hundreds of kcps, depending on the model and operating voltage), which can be significantly reduced by means of a synchronous and proper acquisition of light in a small gate window \cite{sanvito2024assessing}. The integration gate is also crucial to decrease the optical cross-talk effect, that is the probability that, when a cell is fired, also some neighboring cells click \cite{afek,ramilli}. 
While in some models the probability associated to this drawback is natively well below 10$\%$, some experiments may require to operate them with a higher overvoltage to achieve a higher quantum efficiency (up to values larger than 50$\%$), causing an increase in the cross-talk probability. 
Finally, it is also worth mentioning the pile-up effect, that is the probability that two photons simultaneously impinge on the same cell \cite{perina,pritchard}. Its occurrence depends on the size of the light beam with respect to the size and number of cells, specifically, it can be decreased reducing the pixel pitch and increasing the number of cells.\\ 
In our investigation, we consider the models MPPC-S13360-1325CS (25CS hereafter), MPPC-S13360-1350CS (50CS hereafter), and MPPC-S15639-1325PS (25PS hereafter). The 25CS and 50CS have square sensors \SI{1.3}{\milli\meter}$\times$\SI{1.3}{\milli\meter} large, while the 25PS has a rectangular sensor \SI{1.3}{\milli\meter}$\times$\SI{1.1}{\milli\meter} large. The 25CS and 50CS belong to the same series (S13360) and have a maximum quantum efficiency around \SI{450}{\nano\meter} ($ \sim25 \%$ and $ \sim40 \%$, respectively), but have a different pixel pitch (\SI{25}{\micro\meter} and \SI{50}{\micro\meter}, respectively), while the third model has a maximum quantum efficiency ($\sim30\%$) at \SI{660}{\nano\meter} and a pixel pitch equal to \SI{25}{\micro\meter}. Table~\ref{tab:sipm} provides a summary of the relevant features of these detectors.
While the comparison between sensors belonging to the same series allows the investigation of the limits imposed by the pile-up effect, the one between SiPMs with the same pixel pitch enables a study connected to the quantum efficiency.\\
We provide the bias voltage to all the SiPM models using a Caen SP5600 Power Supply and Amplification Unit (PSAU), whose bandwidth range extends up to \SI{500}{\mega\hertz}. 
The results presented in Section~\ref{sec3} involve two acquisition configurations, that will be described in more details in this Section. For the first one (Section~\ref{sec:dynamic range}, excluding synchronous acquisitions), we used \SI{56}{\volt} for both 25CS and 50CS, corresponding to an overvoltage of \SI{4.5}{\volt} and \SI{3}{\volt}, respectively, and \SI{54.9}{\volt} for 25PS, which corresponds to a \SI{10}{\volt} overvoltage. The second (Section~\ref{maxrate} and synchronous acquisitions in Section~\ref{sec:dynamic range}), on the other hand, enabled the measurements over a wider intensity range, therefore we chose to operate the 25CS at \SI{56.5}{\volt}. Indeed, the bias voltage can be tuned within a range to adjust the detector response; in this case it was set to maximize the separation between adjacent photon-number peaks, as shown in Section~\ref{sec:PHS}.
As for the amplification stage, in all cases the gain of the fast amplifier embedded in the PSAU is equal to \SI{17}{\decibel}.
\begin{table}[htbp]
    \centering
    \caption{Main characteristics and operating parameters of the three SiPM
    models employed in this work. Datasheet values for dark count rate, crosstalk, and gain are specified at the recommended operating voltage $V_\mathrm{op} = V_\mathrm{BR} + V_\mathrm{over}$, obtained as the sum of the breakdown voltage and the overvoltage \cite{S13360, S15639}.
    }
    \label{tab:sipm}
    \color{black}
    \begin{tabular}{lccc}
        \toprule
        & \textbf{S13360-1325CS} & \textbf{S13360-1350CS} & \textbf{S15639-1325PS} \\
        \midrule
        Sensor size (\si{\milli\meter\squared})    & $1.3 \times 1.3$  & $1.3 \times 1.3$  & $1.3 \times 1.1$  \\
        Pixel pitch (\si{\micro\meter})            & 25                & 50                & 25                \\
        Number of pixels                           & 2668              & 667               & 2120              \\
        Peak quantum efficiency (\si{\percent})                    & $\sim 25 $ @\SI{450}{\nano\meter}         & $\sim 40$ @\SI{450}{\nano\meter}         & $\sim 30$ @\SI{660}{\nano\meter}         \\
        Bias voltage (\si{\volt})                  & 56                & 56                & 54.9              \\
        Overvoltage (\si{\volt})                   & 4.5               & 3.0               & 10.0              \\
        Dark count rate typ.\ (\si{\kilo\hertz})               & 70                & 90                & 700               \\
        Crosstalk probability (\si{\percent})      & 1                 & 3                 & 4                 \\
        Gain                                       & $7.0\times10^{5}$ & $1.7\times10^{6}$ & $1.3\times10^{6}$ \\
        \bottomrule
    \end{tabular}
\end{table}

\noindent The digitizer employed is the Nuclear Instruments DAQ141, a \SI{1}{\giga\sample\per\second}, \SI{14}{\bit} waveform digitizer evolved from the DAQ121~\cite{Pooley_2023}, with a \SI{2}{\volt} dynamic range, a differential input stage and a sampling period of $T_s = $~\SI{1}{\nano\second}. 
The PSAU suppresses all the signal components above \SI{500}{\mega\hertz}, so that signal aliasing is negligible.
Moreover, the \SI{14}{\bit} resolution over a full scale range of \SI{2}{\volt} reduces the quantization error to \SI{35}{\micro\volt}, which is four orders of magnitude less than the amplitude associated to the detection of a single photon ($\sim$\SI{100}{\milli\volt} after the amplification stages).\\
The Firmware for the DAQ141 is developed using Sci-Compiler~\cite{scicompiler_web}, a graphical environment provided by Nuclear Instruments that allows custom signal-processing pipelines to be deployed on the on-board FPGA without writing hardware description language.\\
The shaped output of the PSAU is single-ended; it is therefore converted to a differential signal by means of an adapter board manufactured by Nuclear Instruments, which provides an additional analog gain of \SI{20}{\deci\bel}. The differential signal is then routed to the digitizer via standard shielded network cables of up to \SI{10}{\meter}, whose common-mode rejection suppresses electromagnetic interference picked up along the cable run; this allows the digitizer to be located far from the optical setup without significant signal degradation. Combined with the configurable digital gain  stages available in the firmware, this allows the overall gain of the acquisition chain to be tuned both to the intensity of the light source and to the specific single-photon amplitude of each SiPM model.\\
In conventional quantum-optics experiments, charge integration is typically performed by analog gated integrators (boxcar averagers, see Appendix~\ref{appendix}). Their main limitation is the acquisition rate they can support, which is below tens of kHz. Moreover, they could be affected by aging effects. Here, the entire analog processing chain downstream of the adapter board is replaced by a digital pipeline running in the DAQ141 FPGA. Since it cannot process analog signals directly, the FPGA is preceded by an ADC \cite{adc}, which is clocked using chips designed specifically for this purpose, so that the intrinsic jitter of the clock is negligible as well as possible temperature drifts. In the following, we refer to the scalar integer values processed by the FPGA as digital numbers.
All operations (baseline subtraction, deconvolution, and charge integration) are carried out on the digitized waveform; the result is therefore fully determined by the register configuration, and identical settings produce identical output independently of environmental conditions. All parameters are accessible at runtime without firmware recompilation.\\
The firmware processes the digitized waveforms through the following chain of blocks, executed in sequence:

\begin{description}

    \item[Baseline Restorer] The signal baseline is continuously estimated as a moving average over a configurable number of samples and subtracted from the input waveform, so that all downstream processing stages operate on a zero-mean signal. When a trigger event is detected, the baseline estimation is inhibited for a configurable hold-off time, preventing the pulse itself from biasing the estimate.
    \item[Pole-Zero Compensation (Digital Deconvolution)] The three-stage amplifier of the PSAU shapes the SiPM signal into a quasi-exponential pulse with decay time constant $\tau$. To recover the original fast charge spike and avoid ballistic deficit in the subsequent integration stage, a digital pole-zero compensation filter is applied. The PSAU output $x[n]$ can be modeled as the convolution of the ideal signal with an exponential impulse response, so that the $z$-domain transfer function of the shaping filter is
    \begin{equation}
        H(z) = \frac{1}{1 - a\,z^{-1}}, \qquad a = e^{-T_s/\tau},
    \end{equation}
    where $T_s = $~\SI{1}{\nano\second} is the sampling period. The inverse (deconvolution) filter is therefore
    \begin{equation}
        H^{-1}(z) = 1 - a\,z^{-1},
    \end{equation}
    which corresponds to the first-order difference equation
    \begin{equation}
        y[n] = G\bigl(x[n] - a\,x[n-1]\bigr),
        \label{eq:pz}
    \end{equation}
    where $G$ is a configurable digital gain factor. 
    The filter is implemented on the FPGA, which processes the multiplexed samples at \SI{250}{\mega\hertz} and produces one output sample per \SI{1}{\nano\second}, achieving a pipeline initiation interval of II$=$1, meaning a new output sample is available at every clock cycle.
    \item[Charge-to-Digital Converter] The deconvolved signal $y[n]$ is integrated over a fixed gate window of duration $T_{\mathrm{gate}} \in [10, 20]~\si{ns}$, yielding a quantity proportional to the collected charge:
    \begin{equation}
        Q = \sum_{n=n_0}^{n_0 + \lfloor T_{\mathrm{gate}}/T_s \rfloor} y[n],
        \label{eq:qdc}
    \end{equation}
    where $n_0$ marks the first sample of the integration window. 
    The firmware implements configurable digital delay lines on both the signal and trigger paths, allowing the position of the integration gate to be adjusted relative to the pulse arrival time and thus achieving an effect equivalent to a pre-trigger acquisition.
    \item[Leading-Edge Discriminator and Trigger Configuration] Acquisition must be synchronized with the periodic light emission of the source. In this work two configurations are employed depending on the source (see Section~\ref{th:light} for details). Specifically, when the sub-picosecond Yb:KGW laser is used, a fraction of each optical pulse is intercepted by a fast photodiode, whose electrical output provides an external trigger signal to the digitizer. When the pulsed diode laser is used instead, the trigger is generated internally by the digitizer and delivered to the laser driver, ensuring that both instruments share the same clock domain and eliminating timing jitter between the trigger and the acquisition window (see Appendix~\ref{appendix sync} for more details). In both cases, a leading-edge threshold discriminator opens the integration gate synchronously with the trigger.
    Operating in this deterministic, synchronous regime, as opposed to a Poissonian one, suppresses the dark-count contribution, since the gate window can be kept sufficiently narrow to exclude thermally generated carriers, whose rate is uncorrelated with the laser clock. For clarity, throughout the remainder of this work the term "synchronous" is reserved for the case of a single clock domain for light emission and detection.
\end{description}

\noindent For each trigger event, the firmware streams the integrated charge values $Q_1$ and $Q_2$ measured by the two SiPMs, reducing the per-event data volume from several hundred samples per channel to a single scalar value per channel and enabling event-by-event reconstruction of the quantum correlations of entangled states (see Section~\ref{th:light}). In addition, the Resource Explorer extension of Sci-Compiler provides live pulse-height spectra and oscilloscope views by reading and writing the digitizer registers directly, allowing the operator to monitor signal quality and adjust acquisition parameters in real time without interrupting the measurement. In particular, the live pulse-height spectra of the two channels provide immediate visual feedback on the mean number of detected photons in each arm, enabling the optical setup to be adjusted until the two arms of the entangled states are balanced.
    
\subsection{The light sources} \label{th:light}
\begin{figure}
\centering
\includegraphics[width=0.9\textwidth]{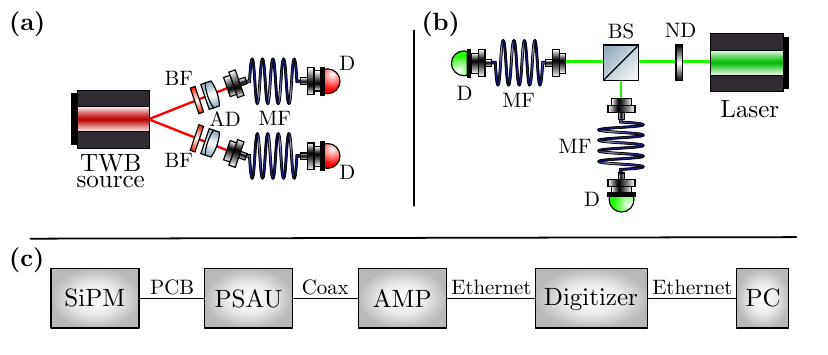}
\caption{Sketch of the experimental setup used to measure TWB states of light (a) or coherent states of light (b). TWB: twin-beam, BF: band pass filter, AD: achromatic doublet, MF: multi-mode optical fiber, D: detection chain, BS: beam splitter, ND: neutral density filter. (c): block diagram of the detection chain. SiPM: silicon photomultiplier, PCB: printed circuit board, PSAU: power supply and amplification unit, Coax: coaxial cable, AMP: amplifier adapter board.}\label{setup}
\end{figure}

\noindent To characterize the acquisition chain we exploit two different kinds of laser sources producing quantum and classical states of light, as shown in Fig.~\ref{setup}(a) and (b), respectively.
This choice is mainly connected to the application fields already described in Section~\ref{sec1}, and specifically to communication protocols based on the reconstruction of photon-number distributions, such as those in Refs.~\cite{razzoli2025hybrid,4states}.  
One of the sources is a sub-picosecond Yb:KGW laser system operating at \SI{3}{\kilo\hertz}. The built-in third harmonic pulses at \SI{343}{\nano\meter}, whose phase fronts are properly tilted by means of a pair of prisms and a de-magnifying telescope, are used to produce multi-mode TWB states through spontaneous parametric down conversion in a $\beta$-barium-borate crystal (\SI{4}{\milli\meter} long, cut angle = 34$^{\circ}$).

\noindent Assuming that the $\mu$ modes are equally populated \cite{machulka}, the generic TWB state can be written as \cite{epl10,silberhorn2016}
\begin{equation} \label{multiTWB}
\vert \Psi^{\mu}_{\rm TWB} \rangle = 
\bigotimes_{k=1}^\mu \sqrt{1- \lambda^2}
\sum_{\nu=0}^{\infty} \lambda^{\nu} \vert \nu\rangle_k \otimes \vert \nu\rangle_k,
\end{equation}
where $k$ labels the modes, $\nu$ is the number of photons in each mode, $\lambda$ is defined through $\lambda^2 = \langle n \rangle / (\mu + \langle n \rangle)$, and $\langle n \rangle$ is the mean total number of photons in either of the two parties of the TWB, with $n=\sum_{k=1}^\mu \nu_k$.\\ 
Two twin portions of the generated light at frequency degeneracy are spectrally and spatially selected at around \SI{40}{\centi\meter} from the crystal, and delivered to the detectors by means of multi-mode optical fibers. The core of the fibers is \SI{1}{\milli\meter} large in order to match the size of the SiPM sensors without losing light and also reduce the probability of the occurrence of the pile-up effect. Each arm of the TWB is characterized by a multi-mode thermal distribution \cite{mandel}:
\begin{equation}\label{multith}
P^{\mu}(n) = \frac{(n+ \mu -1)!}{n!(\mu-1)!(\langle n \rangle/\mu + 1)^{\mu}(\mu/\langle n \rangle +1)^n}.
\end{equation}
Using the first two moments of the distribution, it is also possible to calculate the Fano factor 
\begin{equation} \label{fano}
    F(n) = \frac{\sigma^2(n)}{\langle n \rangle} = \frac{\langle n \rangle}{\mu} + 1,
\end{equation}
$\sigma^2(n)$ being the variance of the distribution.
The state in Eq.~(\ref{multiTWB}) is also characterized by perfect photon-number correlations between the two parties of the TWB in each mode $k$.\\ 
This feature can be quantitatively investigated by calculating the photon-number correlation coefficient
\begin{equation} \label{Gamma}
    \Gamma = \frac{\langle  n_1 n_2 \rangle - \langle n_1 \rangle \langle n_2 \rangle}{\sqrt{\sigma^2(n_1) \sigma^2(n_2)}}.
\end{equation}
However, having a large value of $\Gamma$ is not a sufficient condition for entanglement, as opposed to the criterion based on the noise reduction factor, that is the variance of the photon-number difference between the two parties of TWB normalized to the shot-noise level
\begin{equation} \label{noise_reduction}
    R = \frac{\sigma^2(n_1 - n_2)}{\langle n_1 \rangle + \langle n_2 \rangle},
\end{equation}
where $\langle n_j \rangle$ is the mean number of photons in the $j$-th arm.
Values of $R$ lower than 1 prove that the state in Eq.~(\ref{multiTWB}) is entangled in the number of photons \cite{agliati,pra07}.\\
The quantities in Eqs.~(\ref{multith})-(\ref{noise_reduction}) can also be expressed in terms of quantities that are more accessible from the experimental point of view. In particular, they can be rewritten in terms of detected photons. Assuming Bernoullian detection, the quantities take the same form, except that the number of photons, $n_j$, is replaced by the number of detected photons, $m_j$. In this case, it is possible to demonstrate that for the state in Eq.~(\ref{multiTWB}) the expression of the correlation coefficient $\Gamma$ in Eq.~(\ref{Gamma}) reads as
\begin{equation} \label{Gamma_multith}
    \Gamma = \frac{\sqrt{(\langle m_1 \rangle  \langle m_2 \rangle)/\mu_1 \mu_2} + \sqrt{\eta_1 \eta_2}}{\sqrt{(1+\langle m_1 \rangle / \mu_1)(1+\langle m_2 \rangle / \mu_2)}},
\end{equation}
where $\eta_j$ with $j=1,2$ is the quantum efficiency in the $j$-th arm, whereas that of the noise reduction factor in Eq.~(\ref{noise_reduction}) is
\begin{equation} \label{noise_reduction_multith}
    R = 1- \frac{2 \sqrt{\eta_1 \eta_2}\sqrt{\langle m_1 \rangle \langle m_2 \rangle}}{\langle m_1 \rangle + \langle m_2 \rangle} + \frac{(\langle m_1 \rangle - \langle m_2 \rangle)^2}{\mu (\langle m_1 \rangle + \langle m_2 \rangle)}.
\end{equation}
The second laser source that we use is a pulsed laser diode emitting pulses with a duration of tens of ps at \SI{515}{\nano\meter}. The laser can be externally triggered at a variable repetition rate. In particular, we provide the trigger generated by the digitizer as explained in Section~\ref{th:det}. As for the repetition rates, in this work we considered \SI{500}{\kilo\hertz}, 1, 10 and \SI{20}{\mega\hertz}, in order to investigate the potential of the acquisition system at frequency rates that could be of interest in the field of quantum communication.
As sketched in Fig.~\ref{setup}(b), the laser, after suitable attenuation, is divided at a balanced beam splitter and delivered to the SiPMs by means of two multi-mode fibers with a core of \SI{1}{\milli\meter}.\\
In this case we focus on the generation and measurement of coherent states of light. They are described by a Poissonian distribution in the number of photons
\begin{equation} \label{poiss}
    P(n) = \frac{\langle n \rangle^n}{n!} \exp{\left( -\langle n\rangle \right)},
\end{equation}
whose high-order moments are all equal to the first one, in particular, the variance is equal to the mean value.
\section{Results}\label{sec3}
\subsection{Dynamic range}\label{sec:dynamic range}
\subsubsection{Pulse height spectra}\label{sec:PHS}
\begin{figure}
\centering
\includegraphics[width=0.9\textwidth]{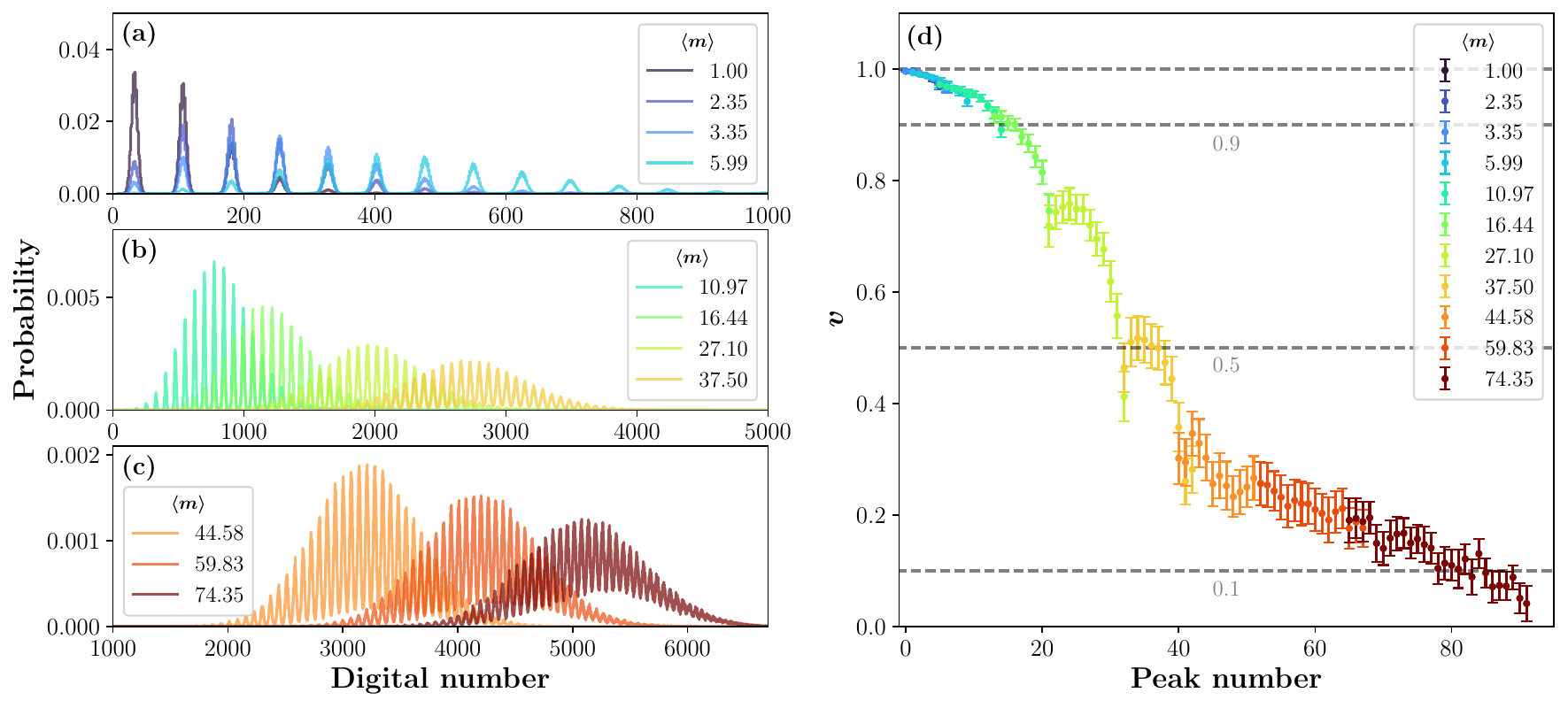}
\caption{(a)-(c): Pulse height spectra as a function of the digital number, resulting from the charge integration, corresponding to coherent states with different mean values (increasing from top to bottom). The employed detector is the model 25CS and the synchronous acquisition is operated at \SI{1}{\mega\hertz}. (d): visibility $v$ as a function of the peak number. The color encoding shows the correspondence between data in the left panels and those in the right panel. The dashed lines correspond to $v = 1, 0.9,0.5,0.1$.}\label{PHS_coh}
\end{figure}

\begin{figure}
\centering
\includegraphics[width=0.9\textwidth]{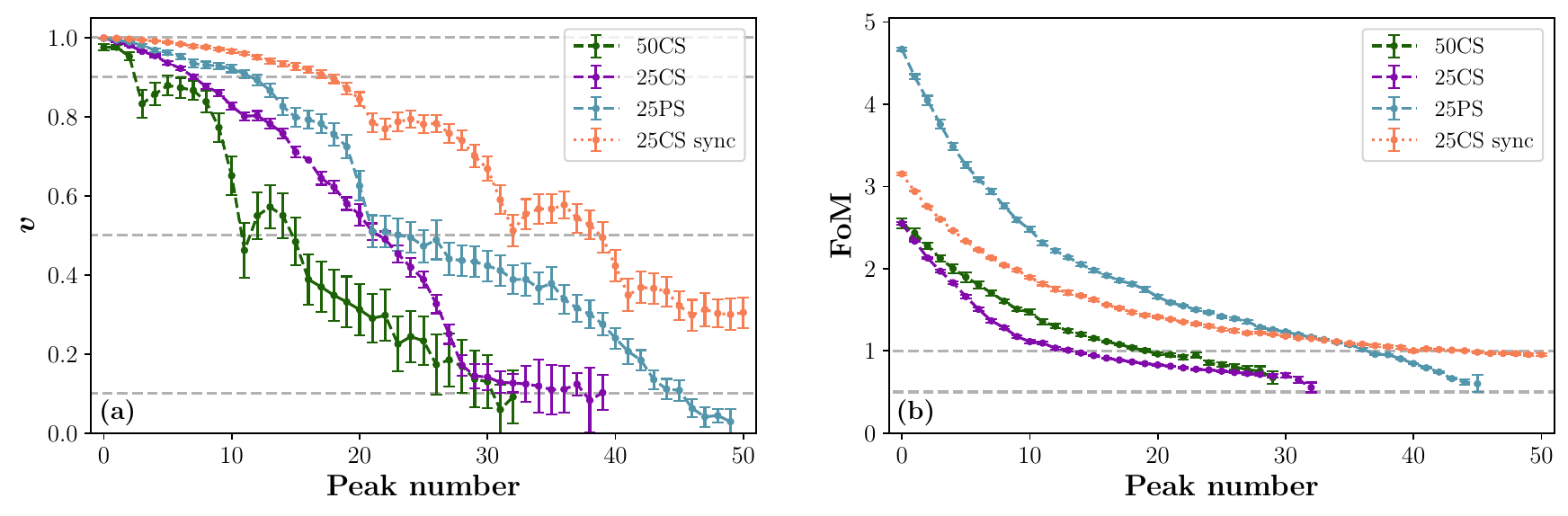}
\caption{Visibility $v$, in panel (a), and figure of merit (FoM), in panel (b), as a function of the peak number for different models of SiPMs operated at \SI{3}{\kilo\hertz} and a synchronous acquisition at \SI{1}{\mega\hertz} with the 25CS. Orange dots in panel (a) correspond to the curve in Fig.~\ref{PHS_coh}(d).
The dashed lines in panel (a) correspond to $v = 1, 0.9,0.5,0.1$, while those in panel (b) to FoM = 1, 0.5. In both panels, lines between dots are used just to guide the eye.}\label{vis_fom}
\end{figure}

\noindent In order to characterize the dynamic range over which the detection chain operates reliably, we consider coherent states of light emitted at \SI{1}{\mega\hertz} and detected by a pair of 25CS operated at a bias voltage of \SI{56.5}{\volt} (i.e. \SI{5}{\volt} overvoltage).
Some examples of pulse-height spectra obtained at different mean values are shown in panels (a), (b), and (c) of Fig.~\ref{PHS_coh}, where the three panels correspond to increasing mean values from top to bottom. 
As it can be noticed, the PNR capability extends to large numbers of photons. However, for peak numbers larger than 80, consecutive peaks have a significant overlap which doesn't allow the correct discrimination. This effect can be preliminary investigated by calculating the visibility of the peaks, which is defined for the $n$-th photon-number peak as 
\begin{equation}
    v_n = \frac{\max_n-\min_n}{\max_n+\min_n},
\end{equation}
where the maximum value, $\max_n$, is calculated as the average of the peak value and its first 6 neighboring points, while the minimum value, $\min_n$, is calculated as the average of the points that lie below half the value of the peak, both on the left and on the right. In Fig.~\ref{PHS_coh}(d) we present the behavior of $v$ as a function of the peak number for the coherent states shown in panels (a)-(c) with the same color encoding, error bars are calculated propagating the errors associated to the mean values of $\max_n$ and $\min_n$. 
The peaks exhibit high visibility ($v>0.9$) up to approximately peak number 15, then $v$ decreases sharply down to 0.3 at approximately peak number 40. For peak numbers above 80, the visibility is very low ($v<0.1$), meaning that peaks cannot be distinguished from the background.\\ 
The evaluation of the visibility allows to perform a first comparison between the different models of SiPM. In Fig.~\ref{vis_fom}(a) we present the visibility as a function of the peak number for the three different models operated with the bias voltages shown in Table~\ref{tab:sipm} (with the exception of the 25CS operated at \SI{56.5}{\volt} for the synchronous acquisition) to detect a portion of TWB at a repetition rate of \SI{3}{\kilo\hertz}. For the sake of clarity, in the same panel we show the visibility already displayed in Fig.~\ref{PHS_coh}(d) as orange dots. 
As it can be clearly observed, the curves corresponding to measurements performed at \SI{3}{\kilo\hertz} decrease more rapidly than the one corresponding to \SI{1}{\mega\hertz}. This different behavior depends on the different laser sources, in particular, as already stated in Section~\ref{th:det}, the laser used to produce the TWB state is not operated in the same clock domain of the digitizer, while the laser used to produce coherent states is triggered directly by the digitizer, therefore it allows a truly synchronous acquisition.
In fact, the asynchronous relationship between the two clocks introduces a timing jitter between the emission of the light pulse and the trigger signal received by the acquisition system, which effectively broadens the uncertainty associated with the integrated signal and reduces the sharpness of the photon-number peaks. \\
Moreover, the visibility exhibited by 25CS and 50CS generally assumes lower values than that corresponding to 25PS, due to the larger gain factor of the latter detector. We also note that the measurements corresponding to 50CS have larger uncertainties because of the lower amount of data at our disposal.\\
To better explore these results, we also consider the figure of merit (FoM). For the n-th photon-number peak, it is defined as
\begin{equation}
    \mathrm{FoM}_n = \frac{\mu_{n+1} - \mu_n}{\mathrm{FWHM}_{n+1} + \mathrm{FWHM}_n} = \frac{\mu_{n+1} - \mu_n}{2\sqrt{2\log2}\,\left(\sigma_{n+1} + \sigma_n\right)},
    \label{eq:fm}
\end{equation}
where $ \mu_n$ and $\sigma_n$ are the mean and standard deviation of the $n$-th Gaussian peak, respectively, obtained from a multi-Gaussian fit of each pulse height spectrum. 
The FoM provides a quantitative criterion for peak discrimination, with values above unity generally indicating resolvable peaks. To give concrete meaning to these numbers, we can calculate the residual overlap between two adjacent Gaussian peaks as a function of FoM. Under the approximation of $\mathrm{FWHM}_1=\mathrm{FWHM}_2$, which is a reasonable approximation for consecutive peaks in the pulse height spectra, the total overlap area between two normalized Gaussian distributions is given by $2\Phi\left(-2\sqrt{2\log2}\cdot\mathrm{FoM}\right)$, where $\Phi$ is the cumulative standard normal distribution. Table~\ref{tab:fom_overlap} shows this relationship quantitatively.\\
This distinction is particularly important when interpreting our results in the context of different measurement approaches. For applications requiring event-by-event discrimination, such as quantum correlation measurements between TWB arms, the FoM directly quantifies the probability of misassigning individual photon-number events, making high FoM values ($\ge 1$) essential to preserve correlations. 
On the other hand, for statistical reconstruction of photon-number distributions via pulse-height histograms, the visibility parameter becomes more relevant, as it characterizes peak prominence above background without requiring perfect event-by-event separation.

\begin{table}[htbp]
\centering
\caption{Residual overlap between adjacent Gaussian peaks as a function of FoM, calculated under the assumption of equal FWHMs. The total overlap represents the fraction of events that would be misclassified when discriminating between two peaks, while the per-peak value indicates the misclassification probability for events originating from a single peak.}
\label{tab:fom_overlap}
    \begin{tabular*}{.7\textwidth}{@{\extracolsep{\fill}}ccc}
        \toprule
        \textbf{FoM} & \textbf{Total overlap (\si{\percent})} & \textbf{Per peak (\si{\percent})} \\
        \midrule
        0.50 & 23.9  & 12.0 \\
        0.75 & 7.74  & 3.87 \\
        1.00 & 1.85  & 0.93 \\
        1.50& 0.041 & 0.021 \\
        2.00  & 0.0002 & 0.0001 \\
        \bottomrule
    \end{tabular*}
\end{table}

\begin{figure}
\centering
\includegraphics[width=0.9\textwidth]{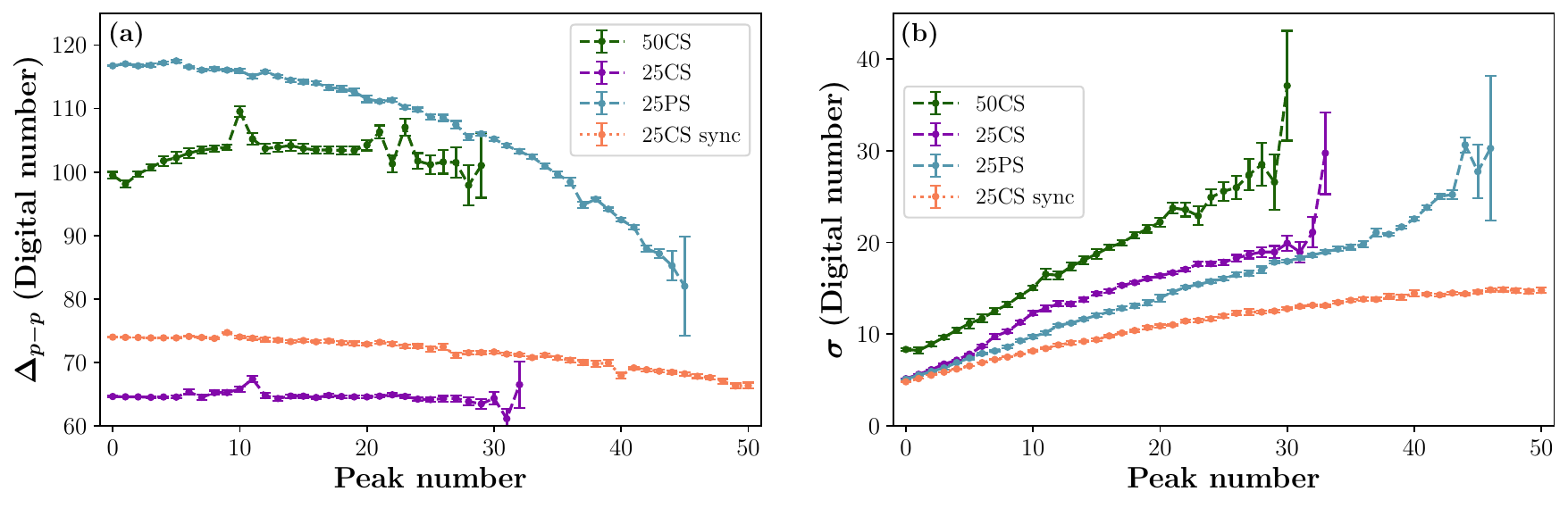}
\caption{Peak-to-peak distance $\Delta_{p-p}$, in panel (a), and standard deviation $\sigma$, in panel (b), in digital numbers as a function of the peak number  for different models of SiPMs operated at 3 kHz and a synchronous acquisition at \SI{1}{\mega\hertz}. In both panels, lines between dots are used just to guide the eye. }
\label{delta_picco}
\end{figure}

\noindent Typical trends of FoM as a function of the peak number are shown in Fig.~\ref{vis_fom}(b), where we compare the different models for the acquisitions at \SI{3}{\kilo\hertz} and also a synchronous one at \SI{1}{\mega\hertz} (a similar analysis performed with an analog system is presented in Appendix~\ref{appendix}). Error bars are calculated by propagating the uncertainties on $\mu_n$ and $\sigma_n$ extracted through the multi-Gaussian fit.
There are two main differences with respect to the results shown in panel (a). Firstly,  for each configuration, the FoM decreases almost monotonically with increasing peak number, as it is independent of the statistics of the incident light. This contrasts with the visibility, which instead exhibits local maxima and minima at the peaks and tails, respectively, of the incident light's photon-number distributions.
Secondly, the highest values of FoM are obtained with the 25PS, which allows perfect discrimination (i.e. FoM$\ge1$) up to peak number 35 (even higher than 2 up to peak number 15). Then the values decrease faster, reaching approximately 0.5 at peak number 45.
Reversely, the 25CS at \SI{1}{\mega\hertz} exhibits lower values at low peak numbers, but there is an inversion at approximately peak number 35 and the FoM remains close to 1 up to peak number 50. Similarly, the 50CS performs better than 25CS at \SI{3}{\kilo\hertz} up to peak number 25, then their values of FoM become comparable. 
As already mentioned, this second difference arises from the definition of $v$ and FoM, since the former evaluates mostly the prominence of the peaks, while the latter involves both the distance of consecutive peaks and their width. 
At lower peak numbers the main parameter that increases the FoM is the distance, while at higher ones the width plays a greater role. This explains why the FoM values at low peak numbers are higher for the detector with the highest gain (i.e. 25PS), while at high peak numbers the best results are obtained using the configuration with the lowest uncertainties (25CS with synchronous acquisition).\\
To explain in a quantitative way the latter statement, in Fig.~\ref{delta_picco} we show the two aforementioned quantities separately for the same data shown in Fig.~\ref{vis_fom}: in panel (a) we plot the peak-to-peak distance, while in panel (b) the standard deviation of the peaks, both as a function of the peak number and expressed in digital numbers. The error bars are obtained directly from the multi-Gaussian fit of the pulse height spectra. The high values of the FoM for the 25PS in Fig.~\ref{vis_fom}(b) are a consequence of the high peak-to-peak distance shown in Fig.~\ref{delta_picco}(a) and relatively low $\sigma$ in Fig.~\ref{delta_picco}(b). In contrast, the 50CS has both high peak-to-peak distance and high $\sigma$, leading to an overall poorer performance.
Concerning the model 25CS, there are two reasons to explain the better performance shown by the measurements taken at \SI{1}{\mega\hertz}. The higher peak-to-peak distance is a consequence of the higher bias voltage, while the synchronous acquisition leads to a lower $\sigma$.\\
Both the peak-to-peak distance and the $\sigma$ corresponding to the measurements taken at \SI{3}{\kilo\hertz} show higher uncertainties at high peak numbers. This behavior can be explained by observing that for those values the visibility is lower than 0.1, meaning that the Gaussian fit is performed only on a small portion of the peak.\\
Finally, we highlight that in panel (a) all the curves show a decreasing behavior at increasing values of the peak number, more evident in the one corresponding to 25PS. 
The nonlinearity in the distance between consecutive peaks is a well-known effect, already studied in different contexts. In some cases, to exclude it, the detection of optical states has been limited to low mean values in order to avoid the onset of the nonlinear behavior \cite{cassina21}. In other cases, linearization methods have been developed to properly reconstruct the statistical properties of the measured states \cite{antonello}.
Here, we highlight that photon-number resolution does not require the peaks to be equally spaced, but only that they can be discriminated. The visibility, as presented above, provides a quantitative criterion for determining the upper limit of the counting range, evaluating the maximum number of peaks that can be identified. Indeed, in the previous figures (see Figs.~\ref{PHS_coh}(d) and \ref{vis_fom}(a)) we have shown that large values of the peak number can be explored. Furthermore, the FoM provides a quantitative metric for comparing the photon-number resolution of different detector models and operating conditions, offering a means to determine the optimal configuration for a given experiment.\\
\subsubsection{Statistical properties}
\begin{figure}
\centering
\includegraphics[width=0.9\textwidth]{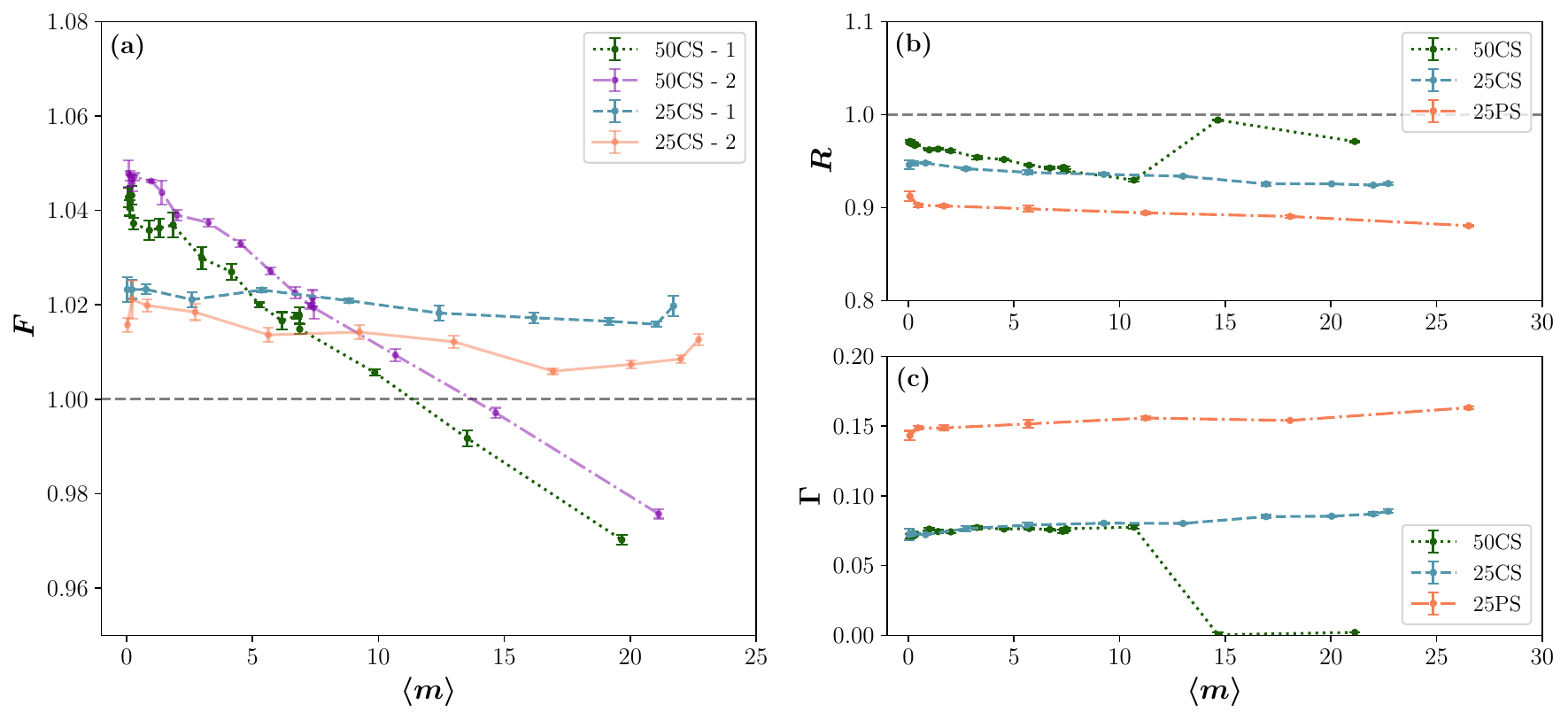}
\caption{(a): Fano factor as a function of the number of photons, $\langle m \rangle$, measured in each TWB arm for the 25CS and 50CS. The curves with similar slope refer to two identical SiPMs, each detecting one arm of the TWB. (b) and (c): $R$ and $\Gamma$, respectively, as a function of $\langle m \rangle$ for the 25CS, 50CS, and 25PS. The lines between dots are used just to guide the eye. In all the panels, error bars are estimated as the standard error of the mean over four repetitions of 250000 acquisitions each.}\label{RFGamma}
\end{figure}

\noindent To deeply investigate the proper reconstruction of the optical states, both classical and quantum, we consider the quantities defined in Section~\ref{th:light}.\\
First of all, we compare the results obtained measuring TWB states with the SiPMs 25CS and 50CS. In Fig.~\ref{RFGamma}(a) we show the Fano factor $F$ of the two twin portions of TWB measured by each detector as a function of the mean number of photons.
It is evident that increasing mean numbers of detected photons, the Fano factor evaluated with 50CS drops below 1, as opposed to theoretical predictions.
Since the values of $F$ corresponding to 25CS are above 1 for all the investigated mean values of the light, the unexpected behavior exhibited by 50CS has to be attributed to the onset of the pile-up effect, which, for a given sensitive area, occurs at lower light intensities in SiPMs with larger pixel pitch, due to the lower number of cells. Despite the use of optical fibers with a core size comparable to the sensor size, the probability that two or more photons impinge on the same cell can still become significant increasing the light intensity. This means that in order to explore larger dynamic ranges, we require the use of SiPMs with smaller pixel pitch. However, as anticipated in Section~\ref{th:det}, a smaller pixel pitch is usually associated with a smaller quantum efficiency. Thus, detecting nonclassical correlations becomes harder.\\
A further proof of the unsuitability of 50CS for detecting light at mean values above $\langle m \rangle \approx 10$ are the plots of $R$ and $\Gamma$, Fig.~\ref{RFGamma}(b) and Fig.~\ref{RFGamma}(c), respectively. In both panels, the behavior of 25CS and 50CS are quite similar up to $\langle m \rangle \approx 10$, then only 25CS gives the expected results, while 50CS experiences a sudden drop in correlation and a consequent rise in the value of $R$. As in the previous example, this behavior originates from the pile-up effect, which is a stochastic process uncorrelated with the TWB.\\
Figure~\ref{RFGamma}(b) and Fig.~\ref{RFGamma}(c) also show the results of the measurements performed with 25PS. The comparison between 25CS and 25PS allows to study the impact of quantum efficiency, since the two models have the same pixel pitch, but the former has a lower quantum efficiency than the latter. Indeed, measurements performed with 25PS show higher values of $\Gamma$ (roughly the double with respect to 25CS), and lower values of $R$. 
Despite the 25PS exhibiting a higher dark count rate and cross-talk compared to the 25CS, the results demonstrate that its higher quantum efficiency enables a more accurate reconstruction of the nonclassical features of quantum states, indicating that the narrow integration gate of the detection system is able to reduce these drawbacks.\\
From these panels it is also possible to investigate the minimum signal that can be correctly detected and reconstructed. Indeed, some of us have stated in previous works that with different kinds of acquisition systems, like analog gated integrators adopted in Ref.~\cite{cassina2025statistical}, the minimum mean value is set to $\langle m \rangle=0.5$ due to the noise of the integrators. In contrast, here we can notice that optical states with mean values as large as 0.07 can be properly reconstructed. This is a crucial point since in quantum communications it is important to have a receiver that is capable of detecting and correctly reconstructing optical states that undergo high loss levels.  
\subsection{Maximum acquisition rate} \label{maxrate}
\noindent For practical applications it is also useful to study the proper operation of the entire detection chain at different repetition and acquisition rates. To this aim, we produce coherent states with different mean values using the pulsed laser diode (see Fig.~\ref{setup}(b)) and we detect them using the 25CS (since they have a higher quantum efficiency at \SI{515}{\nano\meter} compared to 25PS) in synchronous configuration. As already specified in Section~\ref{th:det}, in order to optimize the peak-to-peak distance at large numbers, the 25CS are operated at \SI{56.5}{\volt}.
For the acquisition, we consider four different frequencies, namely \SI{500}{\kilo\hertz}, \SI{1}{\mega\hertz}, \SI{10}{\mega\hertz} and \SI{20}{\mega\hertz}.
It is worth noting that due to the synchronous acquisition, the dark count contribution remains negligible even as the acquisition rate increases. Indeed, performing $10^6$ dark measurements for each condition, we determined a single dark count probability of \SI{0.06}{\percent} and a double dark count probability of \SI{0.0007}{\percent}.
\begin{figure}
\centering
\includegraphics[width=0.9\textwidth]{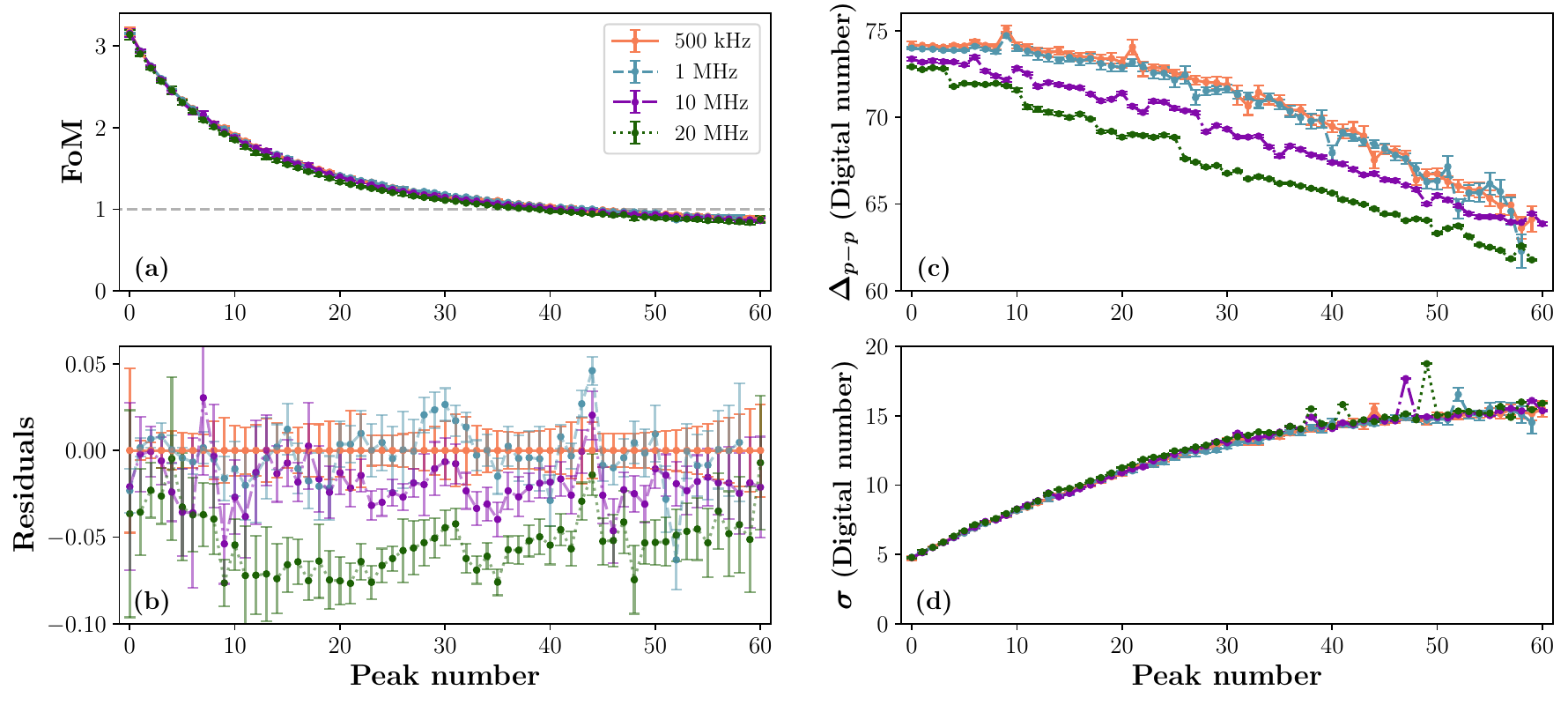}
\caption{(a): FoM as a function of the peak number for the measurements performed with 25CS and a synchronous acquisition at different repetition rates. The gray dashed line corresponds to FoM$=1$. The legend refers to all the panels. (b): Residual of the FoM with respect to the curve corresponding to \SI{500}{\kilo\hertz}. (c) and (d): Peak-to-peak distance, $\Delta_{p-p}$, and standard deviation, $\sigma$, of the peaks, respectively, as a function of the peak number. Error bars are calculated as already explained in the text for Fig.~\ref{vis_fom}(b) and Fig.~\ref{delta_picco}. In all the panels lines between dots are used just to guide the eye.}\label{rep_rate}
\end{figure}

\begin{figure}
    \centering
    \includegraphics[width=0.9\linewidth]{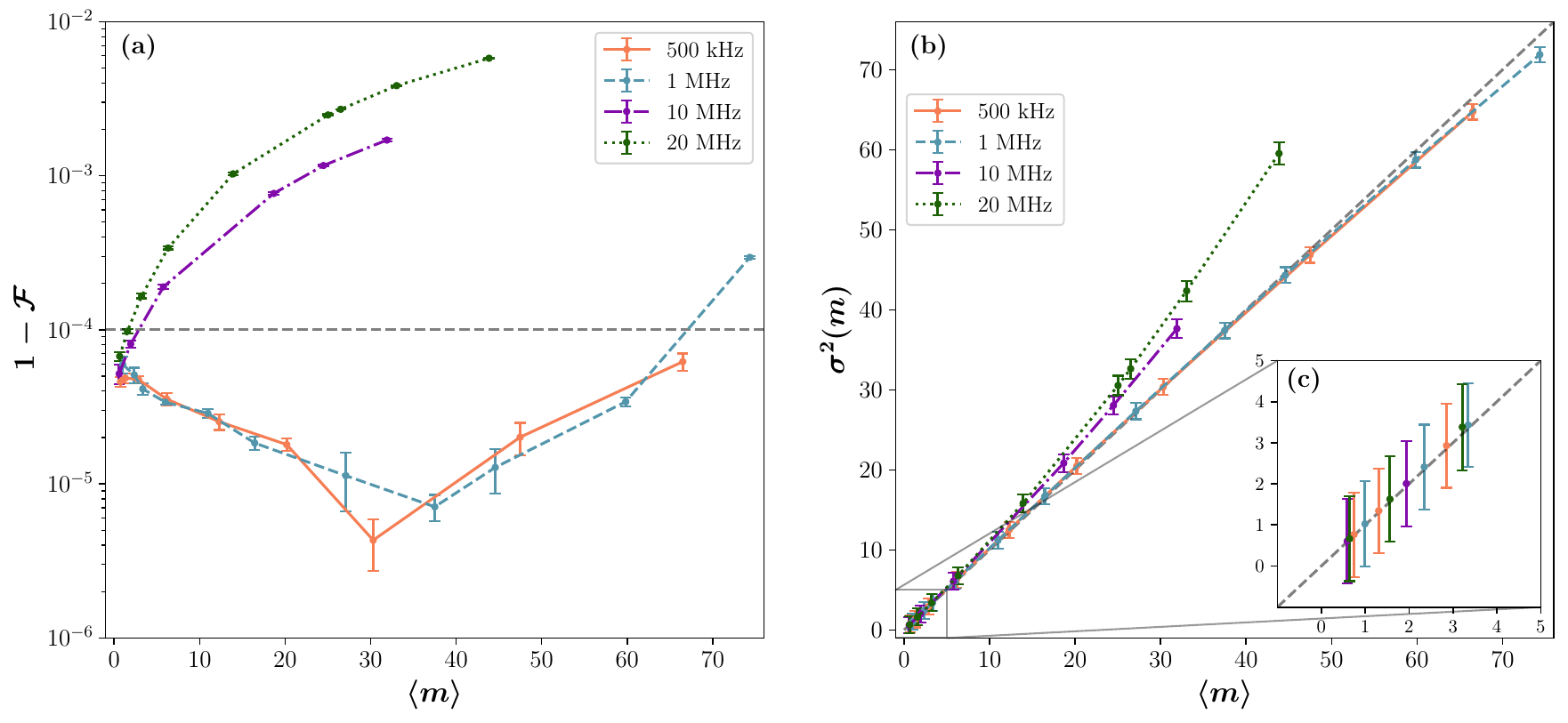}
    \caption{(a): infidelity ($1-\mathcal{F}$) as a function of $\langle m \rangle$ for different acquisition rates. The gray dashed line corresponds to $10^{-4}$. (b): Variance of the reconstructed state as a function of its mean value. The gray dashed line corresponds to the coherent state ($\sigma^2(m) = \langle m \rangle$), the inset (c) shows an enlargement of the region from $\langle m\rangle = 0$ to 5. In both panels, error bars are estimated as the standard error of the mean over four repetitions of 250000 acquisitions each, and lines between dots are used just to guide the eye.} 
    \label{fidelity}
\end{figure}

\noindent In Fig.~\ref{rep_rate}(a) we plot the FoM as a function of the peak number for each condition. All the curves share a common behavior, starting above 3 at peak number 0 and dropping below 1 at approximately peak number 40, while remaining close to 1 up to 50. To better appreciate the differences between the curves, in panel (b) of the same figure we plot the residuals of the FoM with respect to the one corresponding to \SI{500}{\kilo\hertz}. While the residuals corresponding to \SI{1}{\mega\hertz} are randomly distributed around 0, the ones corresponding to \SI{10}{\mega\hertz} and \SI{20}{\mega\hertz} are systematically lower.  
To better understand this difference, in panels (c) and (d) of the same figure we plot the peak-to-peak distance and the $\sigma$ of the peaks, respectively, as a function of the peak number. Concerning the standard deviations, there are no appreciable differences between the different conditions; the peak-to-peak distance, on the other hand, varies significantly across them.
While the results corresponding to lower frequencies (\SI{500}{\kilo\hertz} and \SI{1}{\mega\hertz}) are consistent with each other, the ones corresponding to higher frequencies (\SI{10}{\mega\hertz} and \SI{20}{\mega\hertz}) are systematically lower, resulting in lower values of the FoM. 
We ascribe this behavior to the amplifier stage of the PSAU, since we are employing it beyond its designed scope, requiring both high current and a fast response. This effect becomes even more detrimental at higher intensities of the incident light so that the baseline restorer is no longer able to compensate for the incorrect response of the system. Therefore, each different measurement has been analyzed in post-processing to recover the correspondence between digital number and peak number. Indeed, in order to operate the detection chain correctly at frequencies above \SI{10}{\mega\hertz}, the correct approach should be to use a dedicated amplification stage replacing both the PSAU and the adapter board.\\
To assess the extent to which this effect may be detrimental to the reconstruction of the statistical distribution, in Fig.~\ref{fidelity} we present the infidelity, that is the complementary of the fidelity parameter, defined as $\mathcal{F} = \sum_{m=0}^{\bar{m}}\sqrt{p_{\rm th}(m)p_{\rm exp}(m)}$, where $p_{\rm th}(m)$ and $p_{\rm exp}(m)$ are the theoretical and experimental distributions, respectively, and the sum extends up to the maximum number of detected photons, $\bar{m}$, above which both $p_{\rm th}(m)$ and $p_{\rm exp}(m)$ become negligible. Indeed, the experimental distributions of detected photons have been compared with the theoretical expectations according to Eq.~(\ref{poiss}). The behavior of $1-\mathcal{F}$ is different for the frequencies explored: in the case of \SI{500}{\kilo\hertz} and \SI{1}{\mega\hertz} remains below $10^{-4}$ for high mean values, while in the case of \SI{10}{\mega\hertz} and \SI{20}{\mega\hertz} it rises above $10^{-4}$ even at low mean values. 
Increasing values of infidelity indicate a deviation from the Poissonian distribution. To understand the cause of this deviation, following the analysis presented in Ref.~\cite{pomarico}, in Fig.~\ref{fidelity}(b) we show the plot of the variance of the reconstructed photon-number distributions as a function of the corresponding mean value. The gray dashed line corresponds to the condition $\sigma^2(m)=\langle m\rangle$ (i.e. coherent state), while dots above and below it correspond to super-Poissonian and sub-Poissonian distributions, respectively. The measurements corresponding to \SI{500}{\kilo\hertz} and \SI{1}{\mega\hertz} are compatible with the theoretical expectation up to $\langle m \rangle \approx 60$, then they become lower than expected. As stated in Ref.~\cite{pomarico}, this is a symptom of pile-up effects, since the finite number of cells artificially decreases the variance of the detected photon-number distribution. Instead, a completely different behavior is shown by the measurements corresponding to \SI{10}{\mega\hertz} and \SI{20}{\mega\hertz}, which exhibit a super-Poissonian distribution for $\langle m \rangle >10$. Indeed, increasing at the same time the mean value and the repetition rate, the current requested to the PSAU becomes higher, thus determining an unstable signal output.
This result highlights the presence of a source of excess noise in the system, but while previous results demonstrated that the PSAU is not working correctly at these acquisition rates, further investigations are required to conclude that a different amplifier would also mitigate this problem. 
These results confirm that, while at low frequencies the detection chain can reliably span a wide dynamic range (up to $\langle m \rangle = 60$), at high frequencies this range is considerably reduced.
Finally, in the inset (c) of Fig.~\ref{fidelity} we show an enlargement of panel (b) for low mean values. In this regime, all the measurements are compatible with the Poissonian distribution, meaning that the contribution of cross-talk is completely negligible, otherwise the detected distribution would have been super-Poissonian \cite{chesi19bis, pomarico}.

\section{Discussion}\label{sec_discussion}
\noindent In general, the results shown in Section~\ref{sec3} highlight that the digital acquisition system is properly working and has a good potential for the applications to quantum communication protocols. Compared to systems based on analog charge integrators (e.g. the one used in Refs.~\cite{sanvito2024assessing,cassina2025statistical}), the system presented in this work enables the detection of light at lower intensities and at higher repetition rate. While other detection chains rely on the digital acquisition of the SiPM output \cite{chesi19, chesi19bis, lin, cassina21, pomarico}, to our knowledge none of them supports real-time monitoring of the detected light. Moreover, the digitizers used in Refs.~\cite{chesi19,chesi19bis} have a sampling rate much lower than \SI{1}{\giga\sample\per\second} and induce higher uncertainties on the integrated signal, while the one employed in Ref.~\cite{cassina21} has an acquisition rate limited to approximately \SI{1}{\kilo\hertz}. As opposed to the system presented in Ref.~\cite{lin}, our detection chain is able to mitigate the drawbacks of the SiPMs without the need of cooling the detector and applying a low pass filter to the output signal. Also, increasing the repetition rate from \SI{1}{\mega\hertz} to \SI{20}{\mega\hertz} does not increase the standard deviation of the peaks.\\
In order to obtain the best results, we remark the importance of sharing the same time clock between light emission and detection when working at higher mean values. 
The implementation of a robust phase-locked loop capable of locking to a wide range of frequencies would extend this advantage to laser sources that cannot be directly triggered by the digitizer, thus reducing the jitter well below \SI{1}{\nano\second}. In such configurations, the laser would act as the clock master, with the phase-locked loop enabling the digitizer to synchronize to the laser repetition rate. This approach would ensure that both emission and detection systems operate within the same clock domain, providing the same measurable performance improvements demonstrated in our synchronous measurements, regardless of the laser architecture employed.
A second important aspect concerns the choice of the SiPMs adopted. Indeed, in the case the detection of well-populated states is required, it is important to avoid the occurrence of pile-up effects. This can be obtained by exploiting SiPMs with a smaller pixel pitch, which have a higher number of cells in the same area. 
Another crucial parameter is the quantum efficiency in the employed spectral range. In this regard, we recommend selecting the model with the highest value. The resulting effect of the combination of these considerations can be tested by investigating nonclassical features, such as quantum correlations between the two parties of TWB states.\\
Finally, we point out that the current limit of the detection chain is the maximum repetition rate at which the acquisition can be reliably operated. According to the investigated quantities (peak-to-peak distance, FoM, and infidelity), the chain works as intended up to the repetition rate of \SI{1}{\mega\hertz}, while some problems start occurring at \SI{10}{\mega\hertz}. At this level, the actual limiting factor is the amplification stage provided by the PSAU, whose operation is guaranteed only up to \SI{5}{\mega\hertz} \cite{caen}. A dedicated front-end embedding a faster amplifier would resolve this limitation, enabling acquisitions at repetition rates on the order of tens of MHz.
\section{Conclusions}\label{sec_conclusions}
\noindent In this work, we have investigated the possibility of integrating a digital acquisition system into a detection chain designed for applications in the field of quantum communication.
The device has been tested in combination with different kinds of SiPMs in terms of some quantities that are relevant for the qualification of the entire chain. Since the detectors are endowed with PNR capability, we have investigated the peak-to-peak distance, as well as the visibility and the FoM. The obtained results highlight the crucial role covered by the use of the same time clock to control the laser source and the acquisition system. Indeed, it helps to extend the dynamic range that SiPMs can explore. Reversely, we have shown the limits imposed by the occurrence of pile-up effects by comparing SiPMs with the same sensor area but different pixel pitch. Addressing these two aspects could extend the PNR capability of the detection chain even above 50 photons, a regime that can truly be considered mesoscopic. Moreover, the study of nonclassical correlations has highlighted the impact of the detector's quantum efficiency.
Finally, we have explored the capability of detecting light pulses at high rates. The current chain can properly operate up to \SI{1}{\mega\hertz}, but on-going investigations are now in progress to design a faster amplification stage that could enable higher acquisition rates without compromising the PNR capability of SiPMs and the correct reconstruction of the statistical properties of the optical states.

\appendix
\section{Analog gated integrators}\label{appendix}
\begin{figure}[htp] 
    \centering
    \includegraphics[width=.95\textwidth]{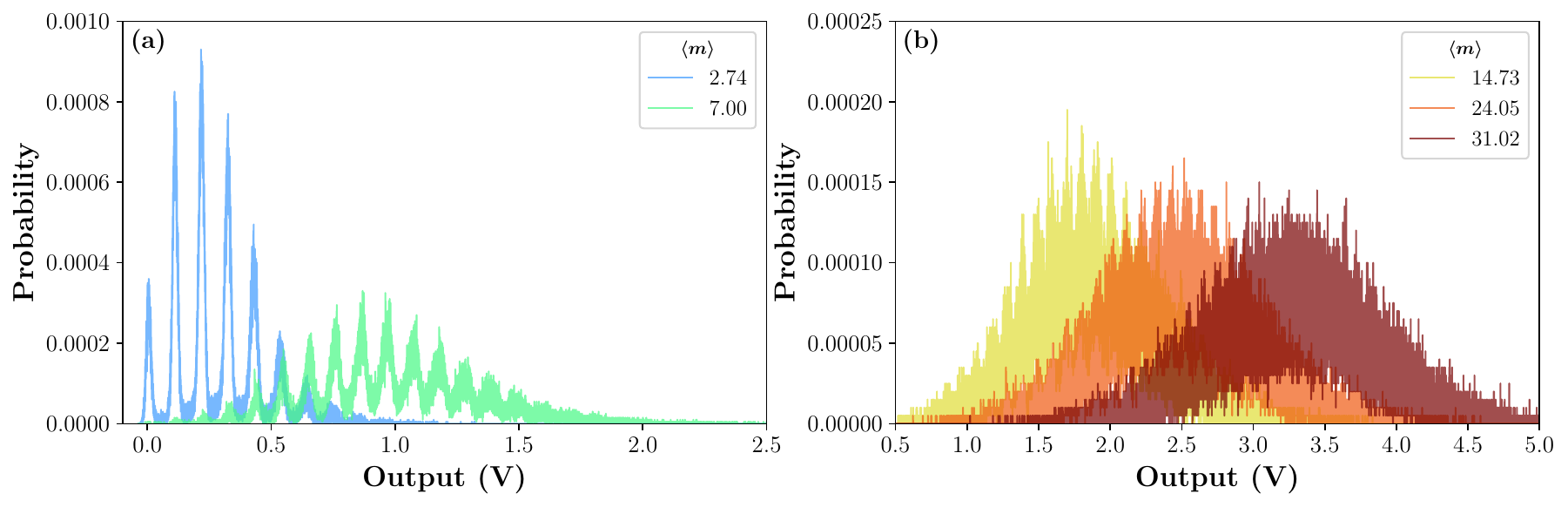}
    \caption{(a)-(b): Pulse height spectra as a function of the output voltages, resulting from analog charge integration, corresponding to coherent states with different mean values. The employed detector is the model 25CS and the acquisition is operated with the analog gated integrators at \SI{3}{\kilo\hertz}.} \label{PHS_old}
\end{figure}
\noindent As already anticipated in the previous Sections, the digital acquisition system presented in this work allowed us to surpass the limitations imposed by analog gated integrators, especially in terms of dynamic range and repetition rate. For the sake of a direct comparison, in this Appendix we show an analysis analogous  to the one presented in Section~\ref{sec3} using the previous acquisition system. 
The detection chain involves the same SiPM detectors presented in the main text (25CS, 50CS and 25PS), and the PSAU operated with the same parameters as in the previous Sections (bias voltage and amplification gain). In this case, the amplified outputs are integrated by means of analog gated integrators (SGI, SR250, Stanford). These boxcars, characterized by a bandwidth of \SI{100}{\mega\Hz}, perform an analog integration of the signal over a gate \SI{10}{\nano\s} narrow around the signal peak. The external trigger coming from the laser source allows their exploitation up to \SI{5}{\kilo\Hz}. For a direct comparison, in the following we show the results obtained at \SI{3}{\kilo\Hz}.
\begin{figure}[htp] 
    \centering
    \includegraphics[width=.9\textwidth]{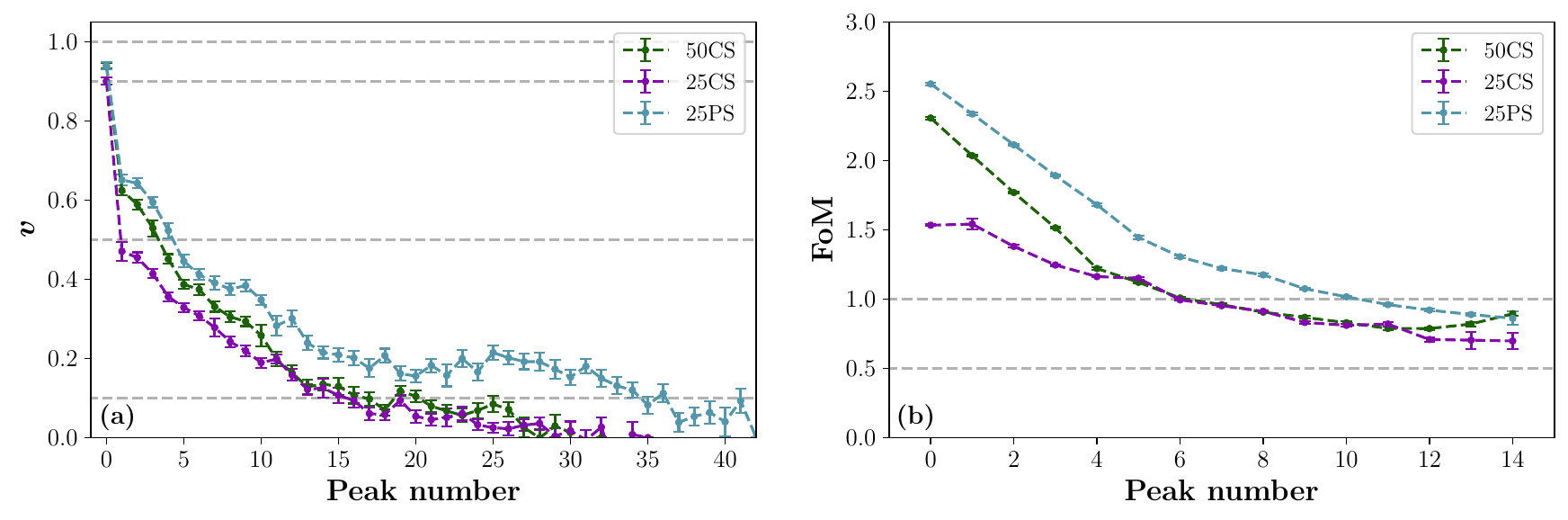}
    \caption{Visibility $v$, in panel (a), and figure of merit (FoM), in panel (b), as a function of the peak number for different models of SiPMs operated at \SI{3}{\kilo\hertz}, using the detection chain based on the analog gated integrators.
    The dashed lines in panel (a) correspond to $v = 1, 0.9,0.5,0.1$, while those in panel (b) to FoM = 1, 0.5. In both panels, lines between dots are used just to guide the eye.} \label{vis_fom_box}
\end{figure}
Finally, the outputs of the boxcars are then digitized by an analog-to-digital converter (ADC, PCI-6251, National Instruments) with a resolution of \SI{16}{\bit} and \SI{20}{\volt} dynamic range, and acquired by a computer.\\
Typical pulse height spectra obtained with this detection chain are shown in Fig.~\ref{PHS_old}. In particular, we present the results corresponding to 25CS for mean values ranging from $\langle m \rangle = 2.7$ to $31$. We remark that while at low mean values (panel (a)), the peaks can be easily distinguished, at higher values (panel (b)) the overlap between consecutive peaks increases rapidly. To quantify the PNR capability, in panels (a) and (b) of Fig.~\ref{vis_fom_box} we plot the visibility and FoM, respectively, for the three SiPM models in the same condition described for Fig.~\ref{vis_fom}. 
Comparing Fig.~\ref{vis_fom_box} and Fig.~\ref{vis_fom}, it is evident that the results obtained in this work are better than the ones achieved using the analog system.
Moreover, we observe that at increasing peak number both quantities decrease more rapidly for the analog system, proving the improvement provided by the digital system in terms of dynamic range. In particular the FoM drops below 1 at peak number 6 (for 25CS and 50CS) and 10 (for 25PS), instead of 15 (for 25CS), 20 (for 50CS) and 40 (for 25PS).
\section{Synchronous acquisition}\label{appendix sync}
\begin{figure}[htp]
    \centering
    \includegraphics[width = .9\textwidth]{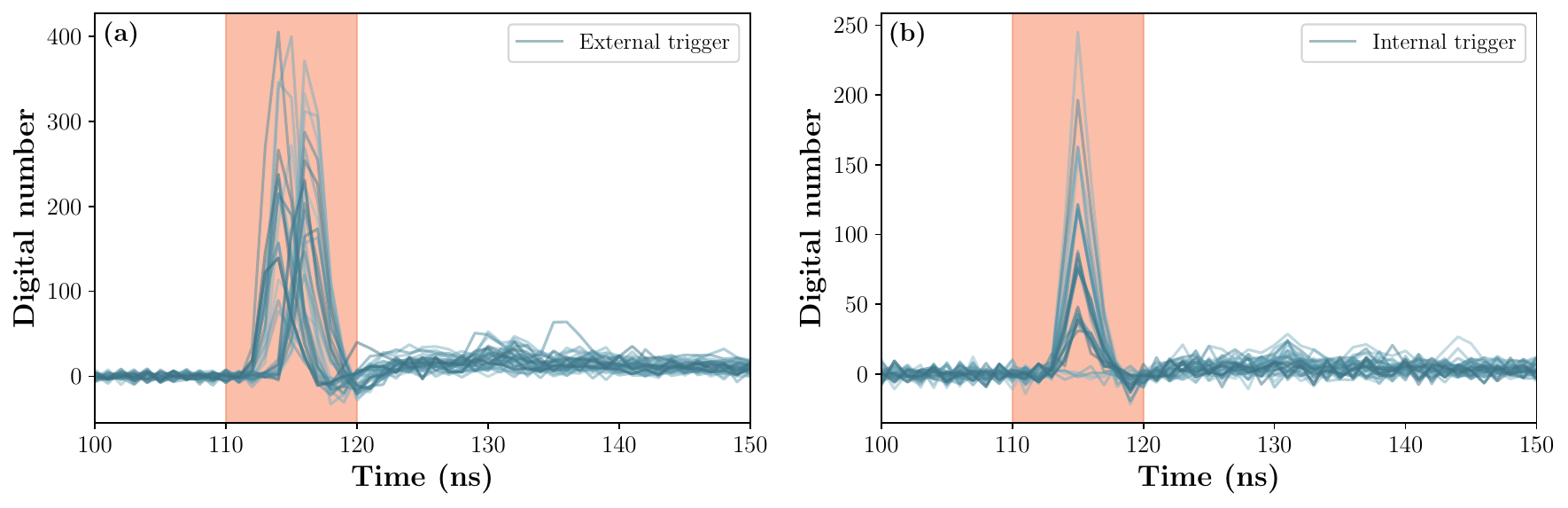}
    \caption{Example of 30 consecutive acquired waveforms (after digital deconvolution) corresponding to measurements of coherent states produced by the diode laser. (a) The laser provides the external trigger to the digitizer; (b) the digitizer drives the laser. The shaded area corresponds to the integration window.}
    \label{fig:jitter}
\end{figure}
\noindent To better highlight the effects of sharing the same clock domain between light emission and acquisition, in Fig.~\ref{fig:jitter} we show some examples of deconvolved waveforms corresponding to measurements performed on coherent states produced by the diode laser and measured with the 25CS biased at \SI{56}{\volt}. Fig.~\ref{fig:jitter}(a) presents the case in which the laser provides an external trigger to the digitizer, hence the two instruments do not share the same clock domain. It can be clearly seen that the different pulses arrive at different times with an uncertainty of about \SI{5}{\nano\second}. Instead, Fig.~\ref{fig:jitter}(b) presents the case in which the digitizer drives laser pulses. Since in this case the two instruments share the same clock domain, the pulses arrival time is determined with an uncertainty below \SI{1}{\nano\second}.

\section*{Acknowledgments}
\noindent We thank Marco Lamperti (University of Insubria) for fruitful discussions.  

\section*{Funding}
\noindent A.~A. acknowledges support from Grant No. PNRR D.D.M.M. 737/2021 and the project “Double weak-field homodyne receiver for
the decoding of quadrature-amplitude-modulated coherent states” supported by University of Insubria. Scientific support from CRIETT centre of University of Insubria (instrument code: MAC27) is greatly acknowledged.

\section*{Abbreviations}
\noindent SiPM, Silicon photomultiplier; PNR, photon-number-resolving; PSAU, Power Supply and Amplification Unit; TWB, twin beam; FoM, figure of merit; 25CS, MPPC-S13360-1325CS; 50CS, MPPC-S13360-1350CS; 25PS, MPPC-S15639-1325PS.

\section*{Availability of data and materials}
\noindent The datasets used and analyzed during the current study are available from the corresponding author on reasonable request.

\section*{Competing interests}
\noindent The authors declare no competing interests.

\section*{Author's contributions}
\noindent AP and AAl conceptualized the work, SC and AAb designed and programmed the digitizer, all the authors performed the measurements, AP and SC analyzed the data, all the authors interpreted them and drafted the work. All the authors have read and approved the final manuscript.

\bibliography{main-bibliography}

\end{document}